\documentclass{PoS}

\usepackage{slashed}
\usepackage{lscape}
\usepackage{array}
\usepackage{graphicx}
\usepackage{amssymb}
\usepackage{textcomp}
\usepackage{amsmath}
\usepackage{latexsym}
\usepackage{subcaption}

\newcommand{\id}{\mathrm{d}}

\title{On the Application of Intersection Theory to Feynman Integrals:
The Univariate Case}

\ShortTitle{Intersection Theory and Feynman Integrals: Univariate}

\author{\speaker{Hjalte Frellesvig}\\
        Niels Bohr International Academy, University of Copenhagen, Blegdamsvej 17, 2100 Copenhagen, Denmark\\
        E-mail: \email{hjalte.frellesvig@nbi.ku.dk}}

\author{Luca Mattiazzi\\
        INFN, Sezione di Padova, Via Marzolo 8, 35131 Padova, Italy\\
        Dipartimento di Fisica e Astronomia, Universit\`a di Padova, Via Marzolo 8, 35131 Padova, Italy\\
        E-mail: \email{luca.mattiazzi@pd.infn.it}}

\abstract{This document is a contribution to the proceedings of the MathemAmplitudes 2019 conference held in December 2019 in Padova, Italy. \\
A key step in modern high energy physics scattering amplitudes computation is to express the latter in terms of a minimal set of Feynman integrals using linear relations.\\
In this work we present an innovative approach (first introduced in~\cite{Mastrolia:2018uzb} and further developed in~\cite{Frellesvig:2019kgj, Frellesvig:2019uqt, Frellesvig:2020qot}) based on intersection theory, in order to achieve this decomposition.
This allows for the direct computation of the reduction, projecting integrals appearing in the scattering amplitudes onto an integral basis in the same fashion as vectors may be projected onto a vector basis.
Specifically, we will derive and discuss few identities between maximally cut Feynman integrals, showing their direct decomposition. \\
This contribution will focus on the univariate part of the story, with the multivariate generalisation being discussed in a different contribution by Gasparotto and Mandal~\cite{ManojProceedings}, and for other contributions see also~\cite{PierpaoloProceedings, Mizera:2020wdt, Weinzierl:2020gda, Brown:2020rda, Vanhove:2020qtt, Bendle:2020iim}.}

\FullConference{MathemAmplitudes 2019: Intersection Theory \& Feynman Integrals\\18-20 December 2019\\	
Padova, Italy}

\begin{document}

\section{Introduction}

\nocite{Mastrolia:2018uzb, Frellesvig:2019kgj, Frellesvig:2019uqt, Frellesvig:2020qot}

When analyzing the enormous amount of data gathered by particle scattering experiments such as those taking place at the Large Hadron Collider at CERN, high precision theoretical predictions based on the Standard Model of particle physics are needed, in order to be able to detect potential tiny discrepancies between theory and experiment, which may hold the key to new physics.

In the language of quantum field theory, precise predictions for the outcome of a particle scattering process require the computation of Feynman integrals, which are mathematical objects given as
\begin{align}
I &= \int \! \frac{\id^d k_i}{\pi^{d/2}} \cdots \! \int \! \frac{\id^d k_L}{\pi^{d/2}} \frac{N (k)}{D_1^{a_1} \! (k) D_2^{a_2} \! (k) \cdots D_{\! P}^{a_{\! P}} \! (k)}
\label{eq:feyndeforig}
\end{align}
where the $D$s are propagators of the form $D_i = (k + p)^2 - m^2$, $d=4 - 2 \epsilon$ is the space-time dimensionality, $k$ and $p$ are $d$-dimensional momenta (internal and external), the $a_i$ are integer propagator powers, $L$ is the number of loops, and $P$ the number of propagators.

Higher precision theoretical predictions require the computation of Feynman integrals with more loops and legs than has been done previously. Adding a loop or a leg to a Feynman integral, implies a huge increase in complexity. For that reason it is important to reduce the set of integrals that has to be computed to a minimum number. Different Feynman integrals in the same family, i.e. integrals differing only by the values of the powers $a_i$, are related through linear relations, allowing for the expression of any integral in a integral family as a linear combination of a minimal set of integrals, known as master integrals
\begin{align}
I &= \sum_{i=1}^{\nu} c_i I_i
\label{eq:masterdec}
\end{align}
with $\nu$ denoting the number of master integrals in the family.

The traditional way of deriving these linear relations, is through the use of integration-by-parts relations~\cite{Chetyrkin:1981qh}, which may be combined in a systematic way using Laporta's algorithm~\cite{Laporta:2001dd}. This algorithm requires the solution of a huge linear system of equations relating the integrals in order to extract the relations needed. For more complicated Feynman integrals this becomes a significant computational challenge and therefore the search for a more direct way to extract relations on the form of eq.~\eqref{eq:masterdec} becomes of higher interest and importance. The approach based on the intersection number, first applied to Feynman integrals in refs.~\cite{Mastrolia:2018uzb} and further developed in~\cite{Frellesvig:2019kgj, Frellesvig:2019uqt, Frellesvig:2020qot} as well as~\cite{Mizera:2019vvs, Weinzierl:2020xyy, Chen:2020uyk}, is a promising piece of progress in that direction.

\section{The loop-by-loop Baikov representation}

Before discussing the intersection theory, let us briefly discuss the loop-by-loop Baikov representation. The Baikov representation~\cite{Baikov:1996iu} (see also~\cite{Lee:2010wea, Grozin:2011mt, Larsen:2015ped, Harley:2017qut, Bosma:2017ens, Frellesvig:2017aai}) for Feynman integrals is a parametric representation, meaning that the integrations are over a number of scalar variables, rather than over the $d$-dimensional loop-momenta of eq.~\eqref{eq:feyndeforig}. Other examples of parametric representations of Feynman integrals are Feynman and Schwinger parametrizations. Baikov representation is characterized by the fact that the integration variables equal the propagators of the original integral. 
A Feynman integral in standard Baikov representation is given by
\begin{align}
I &= \frac{J \; (-i)^L \, \pi^{L-n} \, \mathcal{G}^{(E-d+1)/2}}{\prod_{l=1}^L \Gamma((d{+}1{-}E{-}l)/2)} \int \frac{N(x) \; \mathcal{B}^{(d{-}E{-}L{-}1)/2} \, \id^{n} x }{x_1^{a_1} \cdots x_{\text{P}}^{a_{\text{P}}}}
\label{eq:standard}
\end{align}
Here $L$ is the number of loops, and $E$ the number of independent external momenta.
\begin{align}
n = EL + L(L+1)/2
\label{eq:ndef}
\end{align}
is the number of Baikov variables which again equal the number of scalar products that can be formed between the $L$ loop-momenta and the full set of internal and external momenta.
The two polynomials are
\begin{align}
\mathcal{G} = G(p_1,\ldots,p_E) \;, \quad\qquad \mathcal{B} = G(k_1,\ldots,k_L,p_1,\ldots,p_E)
\end{align}
where $G(y)$ is the Gram determinant of the set $y$ with itself, that is the determinant of the matrix formed by all the scalar products $y_i {\cdot} y_j$. Finally $J$ is the Jacobian of a variable change from the independent scalar products to the Baikov variables, that is $J^{-1} = \text{det} ( d P_i / d (k {\cdot} p)_j )$. The expression for $J$ depends on the exact definition of the propagators, but usually $J = \pm 2^{L-n}$. We see that in most cases $n>\text{P}$, which means that we have to introduce some artificial propagator-like objects to play the role of the remaining Baikov variables.

\begin{figure}[h]
    \centering
   \includegraphics[width=0.37\textwidth]{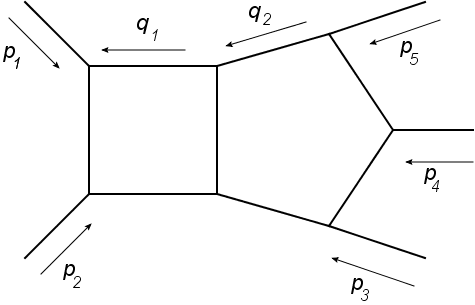}
    \caption{This figure shows the massless pentabox integral, first studied in~\cite{Papadopoulos:2015jft, Gehrmann:2015bfy}. It is a good example for showing the loop-by-loop Baikov parametrizartion. When doing loop-by-loop one has to choose which loop to start from. Starting from the $q_1$-loop-corresponding to $k_1{=}q_2, k_2{=}q_1$ gives the set of propagators containing $k_2$ to be $k_2^2$, $(k_2{+}p_1)^2$, $(k_2{+}p_1{+}p_2)^2$, $(k_2{-}k_1)^2$. From this we see that the set of momenta external to that loop is $\{p_1,p_2,k_1\}$ corresponding to $E_2=3$. The Gram-matrices $\mathcal{B}_2$ and $\mathcal{G}_2$ will contain the scalar products between $k_1 {\cdot} p_1$ and $k_1 {\cdot} p_2$ which together with the remaining propagators show that the $k_1$ integration depends on all four independent external momenta, so $E_1=4$. This gives by eq.~\eqref{eq:ntildedef} $\tilde{n}=2+4+3=9$ showing that only one auxiliary Baikov variable is needed, which we may choose as $x_9=(k_1{+}p_1)^2$.\\
    Alternatively one could have chosen to start from the other side so $k_1{=}q_1, k_2{=}q_2$. In this case the propagators containing $k_2$ would be $k_2^2$, $(k_2{-}p_5)^2$, $(k_2{-}p_5{-}p_4)^2$, $(k_2{-}p_5{-}p_4{-}p_3)^2$, $(k_1{-}k_2)^2$ corresponding to $E_2=4$. This time $\mathcal{B}_2$ and $\mathcal{G}_2$ would introduce the scalar products $k_1{\cdot}p_5$, $k_1 {\cdot} p_4$, and $k_1 {\cdot} p_3$ which together with those coming from the remaining propagators gives $E_1=4$ corresponding to $\tilde{n}=10$ i.e. one more than in the other case. This shows the importance of a clever choice of parametrization of the momenta when applying the loop-by-loop Baikov representation.}
    \label{fig:masslesspentabox}
\end{figure}

The loop-by-loop Baikov representation~\cite{Frellesvig:2017aai} applies the variable changes needed to go to Baikov representation to each loop individually rather than to the whole diagram at once. We define $E_l$ as the number of momenta external to the $l$th loop, after performing the parametrization on the loops labelled with $k_{i>l}$. This number may include the loop-momenta of lower-numbered loops, and also external momenta not directly attached to the $k_l$-loop which may be induced by the previous parametrizations. See fig.~\ref{fig:masslesspentabox} for an example.

In this loop-by-loop case we get the number of integration variables to be
\begin{align}
\tilde{n} = L + \sum_{i=1}^{L} E_i
\label{eq:ntildedef}
\end{align}
which in most cases will be smaller (and never larger) than the number given by eq.~\eqref{eq:ndef}.

In that case, the general loop-by-loop Baikov parametrization of the integral of eq.~\eqref{eq:feyndeforig} is
\begin{align}
I &= \frac{J \; (-i)^L \, \pi^{(L - \tilde{n})/2}}{\prod_{l=1}^L \Gamma((d-E_l)/2)} \int \frac{N (x) \left( \prod_{l=1}^L \mathcal{G}_l^{(E_l-d+1)/2} \, \mathcal{B}_l^{(d-E_l-2)/2} \right) \id^{\tilde{n}} x }{x_1^{a_1} \cdots x_{\text{P}}^{a_{\text{P}}}}
\label{eq:loop-by-loop}
\end{align}
The corresponding polynomials are
\begin{align}
\mathcal{G}_l = G(\{q_l\}) \;, \quad\qquad \mathcal{B}_l = G(k_l,\{q_l\})
\end{align}
where $\{q_l\}$ is the set of momenta external to the $l$th loop. $J$ is defined as above, and is usually given as $J = \pm 2^{L-\tilde{n}}$.

If we take all loops to depend on all lower loop-momenta and all external momenta, we get $E_l = E+l-1$ implying that $\tilde{n}=n$, and it is not hard to see that eq.~\eqref{eq:loop-by-loop} will reduce to eq.~\eqref{eq:standard} in that case.

\section{Theory}

Eq.~\eqref{eq:masterdec} may remind the reader of writing a vector in a vector space in terms of a basis of vectors for that space. And indeed the space of Feynman integrals do live up to all requirements for being a vector space~\cite{Mastrolia:2018uzb}, so referring to the set of master integrals as a basis is not merely a metaphor.



For vector spaces equipped with an inner product, expressing a vector $\langle v |$ in terms of basis vectors $\langle v_i |$ is an easy task, since
\begin{align}
\langle v | = \sum_{ij} \langle v | v_j \rangle \, (\mathbf{C}^{-1})_{ji} \, \langle v_i | \qquad \text{with} \qquad \mathbf{C}_{ij} = \langle v_i | v_j \rangle
\label{eq:decoporig}
\end{align}
and thus
\begin{align}
\langle v | = \sum_i c_i \langle v_i | \quad \Rightarrow \quad c_i = \sum_j \langle v | v_j \rangle \, (\mathbf{C}^{-1})_{ji}
\label{eq:decopusing}
\end{align}

So we realize that the computationally intensive IBP-based algorithm traditionally used to extract the coefficients of eq.~\eqref{eq:masterdec}, may be avoided if we find a way to introduce (what corresponds to) an inner product between Feynman integrals.

To introduce this terminology, we must discuss parametric representations of Feynman integrals, such as those discussed in the previous section. In these representations Feynman integrals are objects of the form
\begin{align}
I &= \int_{\mathcal{C}} u(\mathbf{z}) \varphi(\mathbf{z}) \qquad \text{with} \qquad u(\mathbf{z}) = \prod_i \mathcal{B}_i(\mathbf{z})^{\gamma_i}
\label{eq:feyndef}
\end{align}
where $\mathbf{z}$ is a set of variables, $\mathcal{B}_i$ are polynomials in the $\mathbf{z}$, the $\gamma_i$ are non-integers, and $\varphi$ is a differential form in the $\mathbf{z}$, with a prefactor that is a rational function $\hat{\varphi}(\mathbf{z})$ with the property that all potential poles of $u(\mathbf{z}) \varphi(\mathbf{z})$ are regulated by the $\gamma_i$.

Feynman integrals in the Baikov representation (eqs.~\eqref{eq:standard} and \eqref{eq:loop-by-loop}) are of that form, and so are Feynman integrals in the Lee-Pomeransky form~\cite{Lee:2013hzt} of Feynman parametrization, as long as potential poles in the individual variables are regulated~\cite{Frellesvig:2019kgj} $z_i^{a_i} \rightarrow z_i^{a_i + \rho_i}$ where the regulators $\rho_i$ may be put to zero at the end of the computation. We will focus on the Baikov parametrization in the following.

In addition, also derivatives of Feynman integrals with respect to kinematical variables, and Feynman integrals in shifted dimensions ($d \rightarrow d \pm 2n$) are of the form of eq.~\eqref{eq:feyndef}, so a unified description of integrals of that form will be applicable to all these different objects.

Building on this terminology we may define
\begin{align}
\omega &= d \log( u )
\end{align}
to be of use later. This object has the property that the number of master integrals $\nu$ may be found as~\cite{Lee:2013hzt}
\begin{align}
\nu \;\; = \;\; \text{number of solutions to ``$\omega = 0$''}  
\end{align}
something that will be used extensively in the following.

The integral of eq.~\eqref{eq:feyndef} may be reexpressed as a pairing between a \textit{twisted co-cycle} $\langle \varphi |$ and a \textit{twisted cycle} $| \mathcal{C} ]$
\begin{align}
I \,=\, \int_{\mathcal{C}} \! u(\mathbf{z}) \varphi(\mathbf{z}) \,=\, \langle \varphi | \mathcal{C} ]
\end{align}
where the multi-valued function $u$ is implicit in the definition of the pairing. The $\langle \varphi |$ and $| \mathcal{C} ]$ should be understood not as being necessarily equal to the $\varphi$ and $\mathcal{C}$ of eq.~\eqref{eq:feyndef}, but rather as representatives of the group of forms and integration contours respectively, which integrate to the same $I$.

It is likewise possible to define a \textit{dual} twisted cycle and cocycle as $\int_{\mathcal{C}} u^{\! -1} \! (\mathbf{z}) \varphi(\mathbf{z}) = [ \mathcal{C} | \varphi \rangle$. This allows for the introduction of pairings between the cycles and the co-cycles themselves $\langle \varphi_L | \varphi_R \rangle$ and $[ \mathcal{C}_L | \mathcal{C}_R ]$ which are known as \textit{intersection numbers}. For the complete definition of the intersection numbers, we must refer to the mathematical literature~\cite{cho1995, matsumoto1998}. In the following we will focus solely on the intersection number of the twisted co-cycles $\langle \varphi_L | \varphi_R \rangle$, which play the role of the inner product we were searching for.

With this definition in place, we now have a way of extracting the master integral coefficients of eq.~\eqref{eq:masterdec}. This can be done~\cite{Mastrolia:2018uzb} using eq.~\eqref{eq:decopusing} as
\begin{align}
\langle \varphi | \mathcal{C} ] = \sum_i c_i \langle e_i | \mathcal{C} ] \quad \Rightarrow \quad c_i = \sum_j \langle \varphi_L | h_j \rangle (\mathbf{C}^{-1})_{ji}
\label{eq:decomp}
\end{align}
with $\mathbf{C}_{ij} = \langle e_i | h_j \rangle $ as in eq.~\eqref{eq:decoporig} and with $\langle e_i |$ and $| h_i \rangle$ being the bases of cocycles and dual cocycles respecitively.

\section{Univariate intersection numbers}

Let us start by the univariate case in which the integral of eq.~\eqref{eq:feyndef} is with respect to just one variable $z$. For that case the intersection number is given as
\begin{align}
\langle \varphi_L | \varphi_R \rangle &= \sum_{p \in \mathcal{P}} \text{Res}_{z=p} ( \psi_p \varphi_R )
\label{eq:intdefuni}
\end{align}
with
\begin{align}
(d + \omega) \psi_p = \varphi_L
\label{eq:difeq}
\end{align}
The sum goes over $\mathcal{P}$, the set of poles of $\omega$, and $\psi$ is defined as the solution to eq.~\eqref{eq:difeq}, but as indicated by the subscript all that is needed for practical applications is a solution that is valid around the point $p$, as the residue operation is local, so the defining equation can be solved using a series ansatz.

Let us illustrate this formula and eq.~\eqref{eq:decomp} on the Gauss' hypergeometric function
\begin{align}
{}_2 F_1 (a,b,c;x) &= \int_0^1 \!\! u(z) \, \id z \qquad \text{with} \qquad u(z) = \frac{\Gamma(c)}{\Gamma(b) \Gamma(c-b)} z^{b-1} (1-z)^{c-b-1} (1 - xz)^{-a}
\label{eq:2F1def}
\end{align}
This corresponds to eq.~\eqref{eq:feyndef} with $\mathcal{C} = [0,1]$ and $\varphi = 1 \id z$.

Hypergeometric functions are related through relations called \textit{contiguity relations}, which relate different ${}_2 F_1$ for which the indices $a,b,c$ differ by integers. This is the equivalent of the linear relations relating Feynman integrals in the same family. One example of a contiguity relation is
\begin{align}
{}_2 F_1 (a,b{+}1,c{+}1;x) &= c_1 \, {}_2 F_1 (a,b{+}1,c;x)  + c_2 \, {}_2 F_1 (a,b,c;x)
\label{eq:contiguity}
\end{align}
The values of $c_1$ and $c_2$ are known, but we will try to derive them using the intersection-based formalism. With the $u$ given above, these three integrals correspond to
\begin{align}
\hat{\varphi} = \frac{cz}{b}\;, \qquad \hat{e}_1 = \frac{(1+b-c)z}{b(z-1)}\;, \qquad \hat{e}_2 = 1\;,
\label{eq:Fphilist}
\end{align}
with $\varphi = \hat{\varphi}\,  \id z$ and $e_i = \hat{e}_i \id z$.

The first step is to compute $\omega$:
\begin{align}
\omega = \id \log(u) &= \left( \frac{b-1}{z} + \frac{b-c+1}{z-1} + \frac{ax}{1-zx} \right) \id z
\end{align}
Solving for $\omega=0$ gives two solutions corresponding to the $\nu=2$ master integrals appearing on the RHS of eq.~\eqref{eq:contiguity}.

In order to be able to use the projecting formula eq.~\eqref{eq:decomp}, we also need a set of dual forms $|h_i \rangle$. They can be chosen almost arbitrarily as long as no poles are present that are not poles of $\omega$. We chose
\begin{align}
\hat{h}_1 = \frac{1}{z} - \frac{1}{z-1}\;,\quad \qquad \hat{h}_2 = \frac{1}{z} - \frac{x}{xz-1}\;.
\label{eq:Fphistarlist}
\end{align}
To apply eq.~\eqref{eq:decomp} we need to compute all intersection numbers between the forms of eqs.~\eqref{eq:Fphilist} and \eqref{eq:Fphistarlist}, i.e.:
\begin{align}
\Big\{ \langle \varphi | h_1 \rangle, \langle \varphi | h_2 \rangle, \langle e_1 | h_1 \rangle, \langle e_1 | h_2 \rangle, \langle e_2 | h_1 \rangle, \langle e_2 | h_2 \rangle \Big\}
\end{align}
Let us focus on $\langle e_2 | h_1 \rangle$. To find the intersection number we need to perform the sum of eq.~\eqref{eq:intdefuni}. It goes over the set of poles of $\omega$ i.e.
\begin{align}
\mathcal{P} &= \left\{ 0, \, 1, \, \tfrac{1}{x}, \, \infty \right\}
\end{align}
Let us start by the contribution from $z=0$. In that point we need to find a solution $\psi_0$ to eq.~\eqref{eq:difeq} on the form of a series ansatz, that is
\begin{align}
\psi_0 = \sum_i \, \kappa_i z^i
\end{align}
solving eq.~\eqref{eq:difeq} order by order in $z$, we find the solution
\begin{align}
\kappa_{\leq 0} = 0 \,,\quad \kappa_{1} = \frac{1}{b} \,, \quad \kappa_{2} = \frac{c-b-1-ax}{b(b+1)} \,,\quad \ldots
\end{align}
and inserting that solution into the residue of eq.~\eqref{eq:intdefuni} gives
\begin{align}
\text{Res}_{z=0} ( \psi_0 h_1 ) &= 0
\end{align}
The residues in $z=1$ and $z=\frac{1}{x}$ are likewise $0$, only the pole in infinity contributes with the value $\frac{1}{c-a-1}$, and therefore
\begin{align}
\langle e_2 | h_1 \rangle &= \sum_{p \in \mathcal{P}} \text{Res}_{z=p} ( \psi_p h_1 ) = \frac{1}{c-a-1}.
\end{align}
The same computation can be done for the other intersection numbers, with the results
\begin{align}
\langle \varphi | h_1 \rangle &= \frac{c (a - a x + b x)}{b (c - a) (c - a - 1) x}\,,   &  \langle \varphi | h_2 \rangle &= \frac{c (c + x + b x - c x - 1)}{b (c - a) (c - a - 1) x^2}\,, \nonumber \\
\langle e_1 | h_1 \rangle &= \frac{b - a}{b (c - a - 1)}\,,                       & \langle e_1 | h_2 \rangle &= \frac{1 + b - c}{b (c - a - 1) x}\,, \\
\langle e_2 | h_1 \rangle &= \frac{1}{c - a - 1}\,,                               & \langle e_2 | h_2 \rangle &= \frac{1}{(c - a - 1) x}\,. \nonumber
\end{align}

Inserting these results into the projection formula eq.~\eqref{eq:decomp}, we get the results
\begin{align}
c_1 = \frac{c(1-x)}{x(a-c)} \;,\qquad c_2 = \frac{-c}{x(a-c)} \;,
\end{align}
which makes the contiguity relation of eq.~\eqref{eq:contiguity} true, something that can be checked with the literature or numerically.

The same approach can be applied to the most general univariate generalization of the ${}_2 F_1$, known as the Lauricella $F_D$ function~\cite{Keiji-MATSUMOTO2013367} defined as
\begin{align}
& \;\;\;\;\; F_D (a,b_1,\ldots,b_n,c;x_1,\ldots,x_n) \nonumber \\
&= \frac{\Gamma(c)}{\Gamma(a) \Gamma(c-a)} \int_0^1 \!\! z^{a-1} (1-z)^{c-a-1} \prod_i^n (1 - x_i z)^{-b_i} \id z \,.
\label{eq:lauricella}
\end{align}
This integral family has $n+1$ master integrals, the coefficients of which can be extracted using eq.~\eqref{eq:decomp} as above.

\section{The (maximal cut of the) double box}

\begin{figure}[h]
    \centering
   \includegraphics[width=0.35\textwidth]{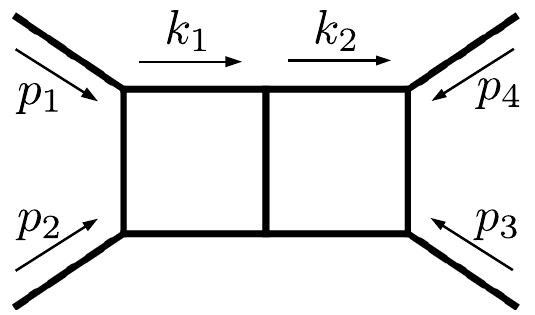}
    \caption{The massless doublebox}
    \label{fig:masslessbox}
\end{figure}
After the purely mathematical example of the ${}_2 F_1$ let us try (following ref.~\cite{Frellesvig:2019kgj}) an example relevant for real life particle physics, the double box seen on fig.~\ref{fig:masslessbox}. This is a Feynman integral on the form of eq.~\eqref{eq:feyndeforig}, and it is given in terms of the propagators
\begin{align}
    D_1 = k_1^2\,, \;\;\; D_2 = (k_1-p_1)^2\,, \;\;\; D_3 = (k_1-p_1-p_2)^2\,, \;\;\; D_4 = (k_1-k_2)^2\,, \nonumber \\
    D_5 = (k_2-p_1-p_2)^2\,,\;\;\; D_6 = (k_2+p_4)^2\,,\;\;\; D_7 = k_2^2\,,\;\;\; z = (k_2-p_1)^2\,.
\end{align}
where the $z$ represents an irreducible scalar product (ISP).

We will analyze it on the maximal cut, in which case the loop-by-loop version~\cite{Frellesvig:2017aai} of Baikov representation given by eq.~\eqref{eq:loop-by-loop} allows for a univariate representation:
\begin{align}
I_{\text{dbox}} &= \int_{\mathcal{C}} \! u \, \id z \qquad \text{with} \qquad u \propto z^{\frac{d}{2}-3} (z+s)^{2-\frac{d}{2}} (z-t)^{d-5}
\label{eq:dboxbaik}
\end{align}
We see that this integral is equivalent to the ${_2} F_1$ defined in eq.~\eqref{eq:2F1def} up to a variable change, and therefore the rest of the computation will proceed along similar lines.

Let us say we want to reduce the integral having two powers in the numerator of the ISP corresponding to the variable $z$, to a basis of master integrals with zero and one powers respectively. That is
\begin{align}
I_{1111111;-2} = c_0 I_{1111111;0} + c_1 I_{1111111;-1} \,,
\end{align}
corresponding to the co-cycles $\varphi =  z^2 \id z$, $\varphi_0 = \id z$, and $\varphi_1 = z \id z$. 
Going through the same procedure as above, tells us that for using eq.~\eqref{eq:decomp} we need to compute six intersection numbers, those of these three co-cycles with a set of dual co-cycles. Computing those intersection numbers and combining the according to eq.~\eqref{eq:decomp} gives the results
\begin{align}
c_0 = \frac{(d-4)st}{2(d-3)} \;, \qquad c_1 = \frac{2t-3(d-4)s}{2(d-3)} \;,
\end{align}
in agreement with the results obtained in the literature and by standard IBP-based codes such as FIRE~\cite{Smirnov:2014hma}.

\begin{figure}
\centering
\begin{subfigure}{0.19\textwidth} \includegraphics[scale=0.19]{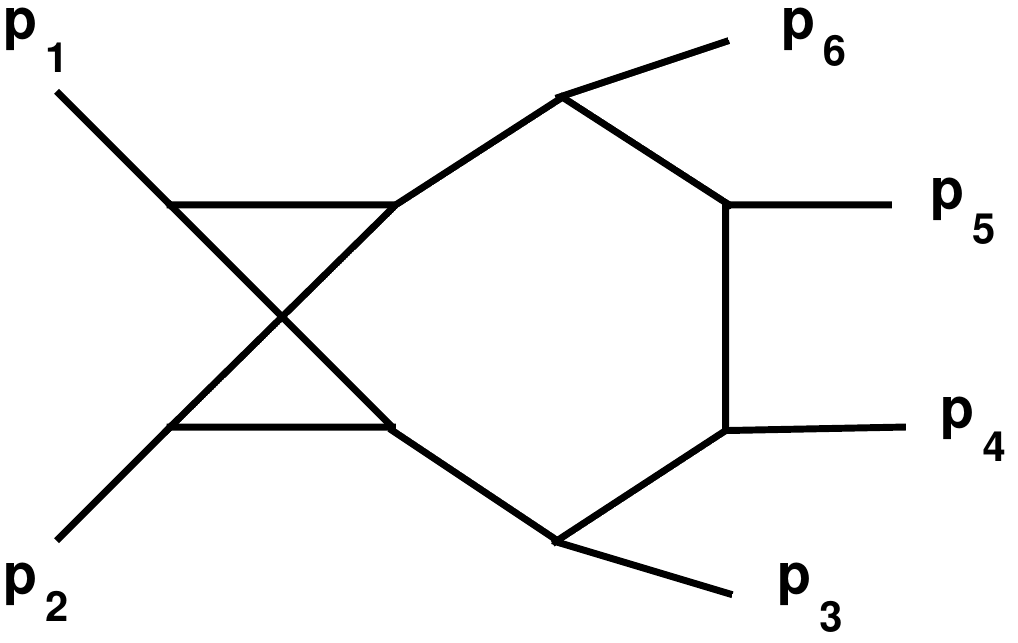} \end{subfigure}
\begin{subfigure}{0.19\textwidth} \includegraphics[scale=0.19]{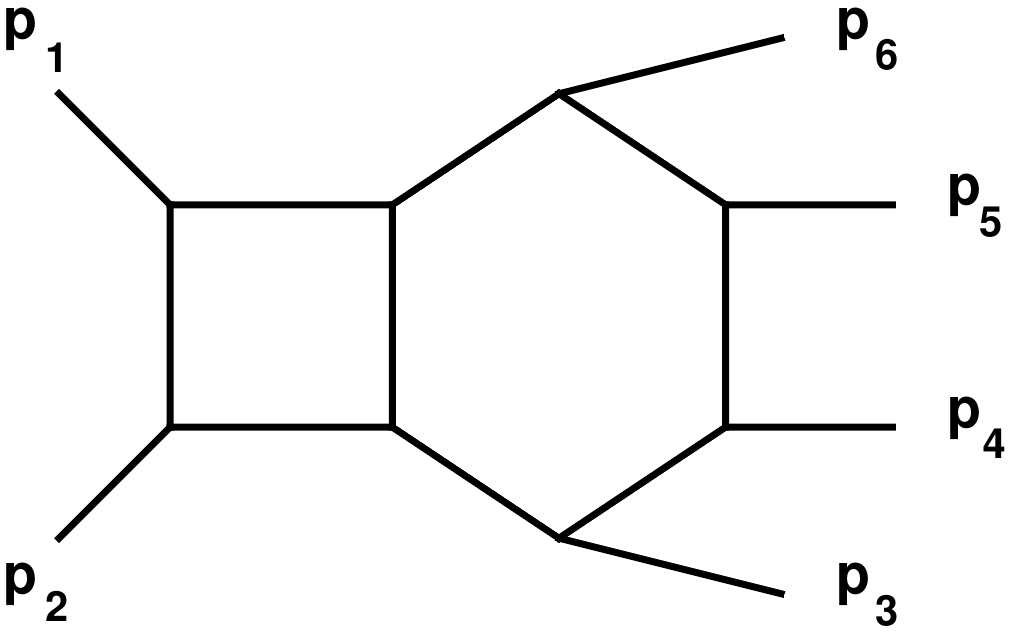} \end{subfigure}
\begin{subfigure}{0.19\textwidth} \includegraphics[scale=0.19]{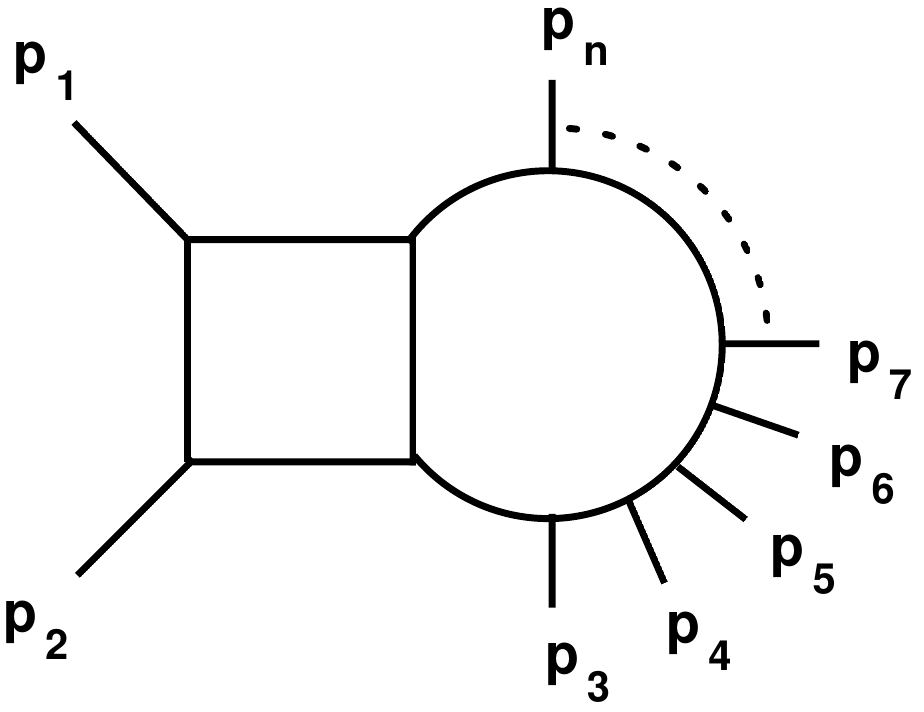} \end{subfigure}
\begin{subfigure}{0.19\textwidth} \includegraphics[scale=0.19]{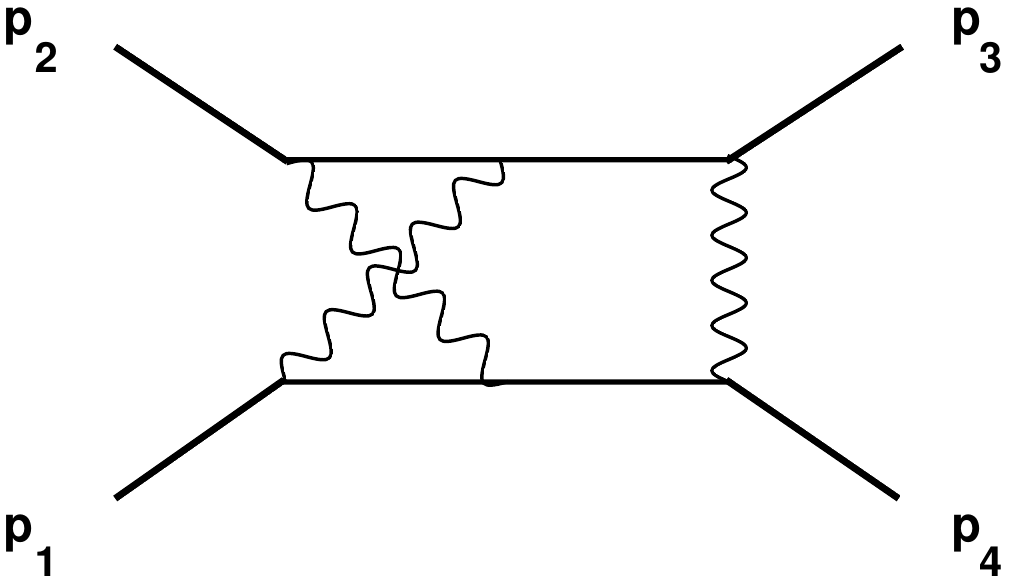} \end{subfigure}
\begin{subfigure}{0.19\textwidth} \includegraphics[scale=0.19]{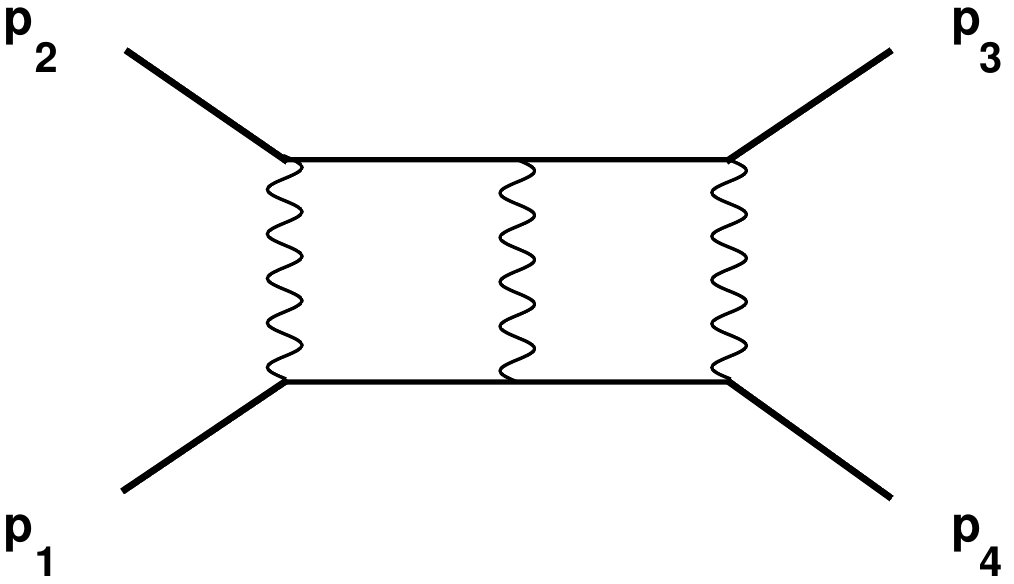} \end{subfigure} \\[1mm]
\begin{subfigure}{0.19\textwidth} \includegraphics[scale=0.19]{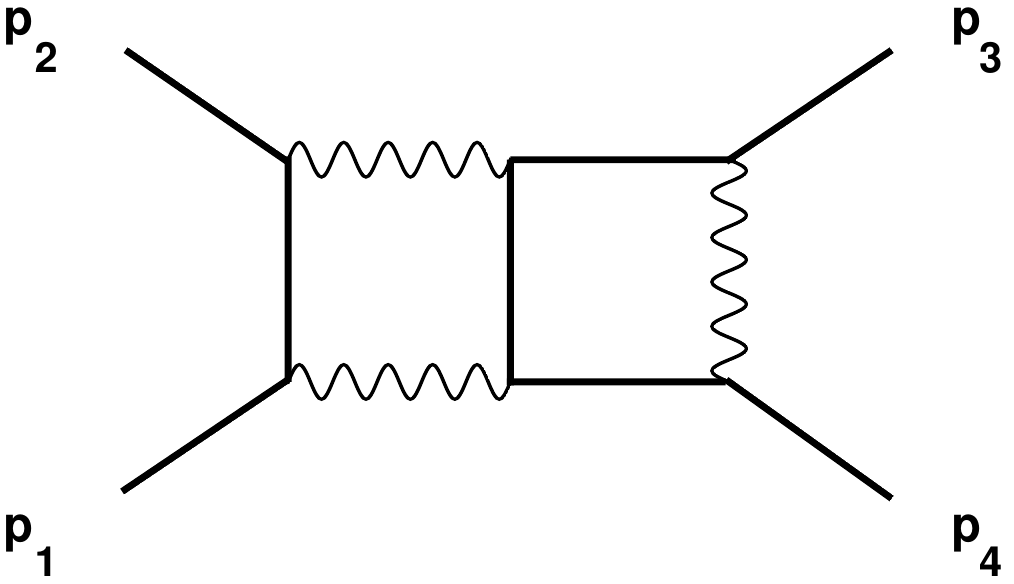} \end{subfigure}
\begin{subfigure}{0.19\textwidth} \includegraphics[scale=0.19]{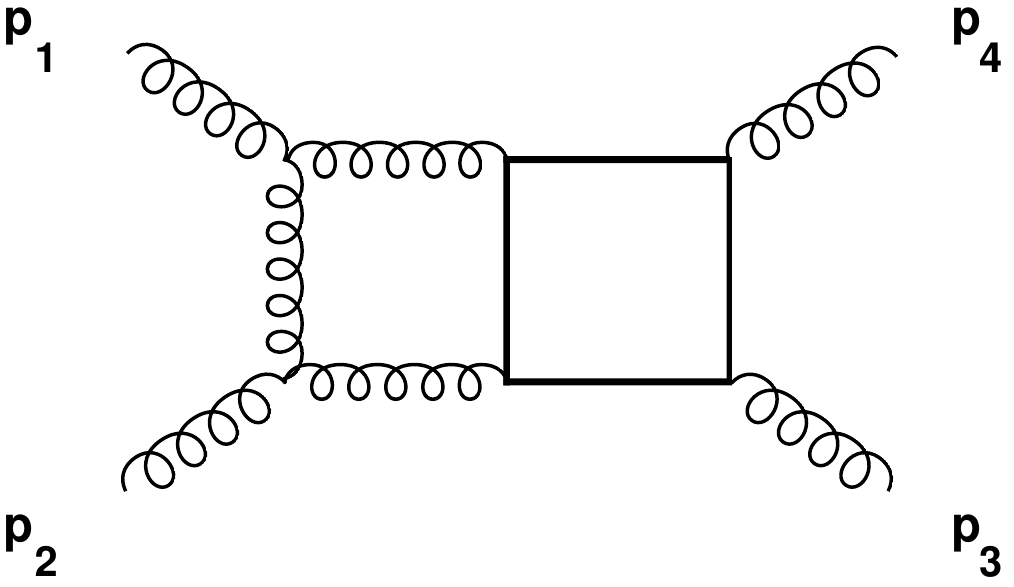} \end{subfigure}
\begin{subfigure}{0.19\textwidth} \hspace{2.3mm} \includegraphics[scale=0.19]{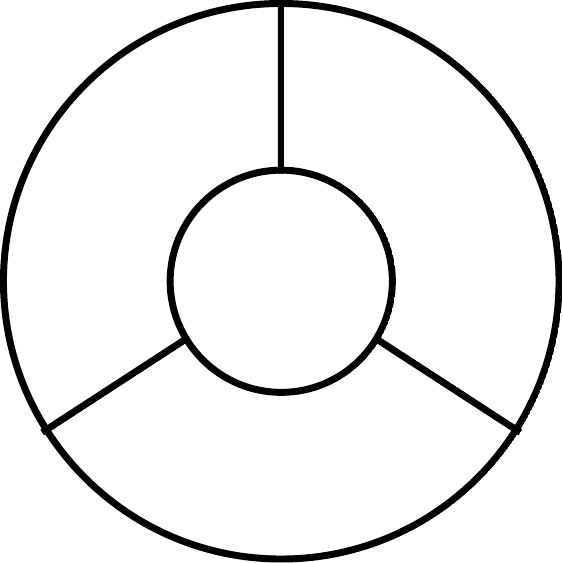} \end{subfigure}
\begin{subfigure}{0.19\textwidth} \includegraphics[scale=0.19]{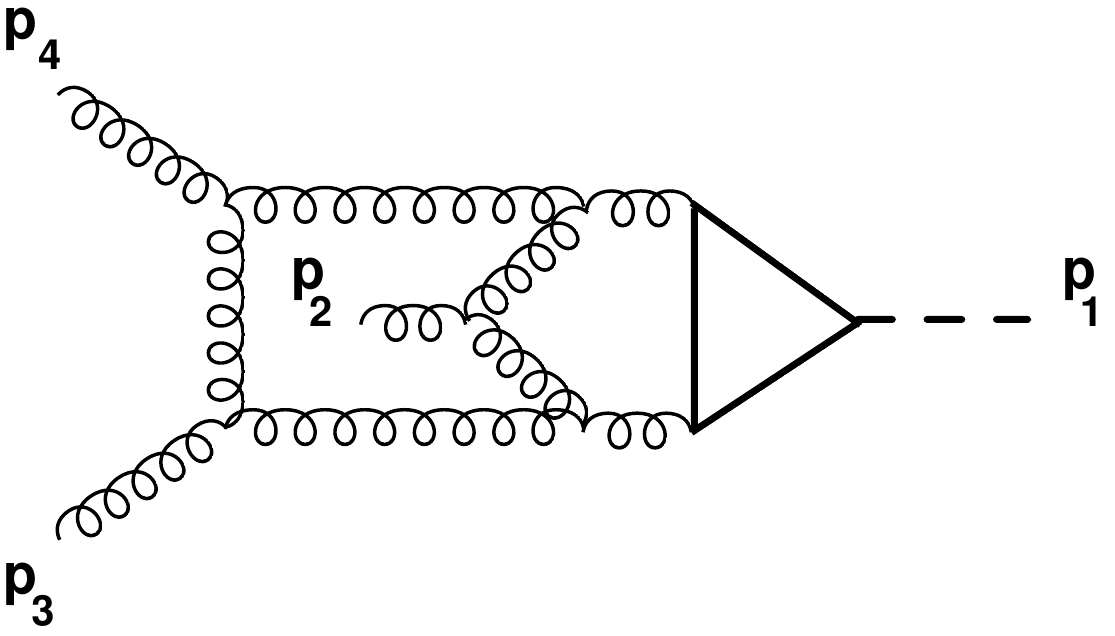} \end{subfigure}
\begin{subfigure}{0.19\textwidth} \includegraphics[scale=0.19]{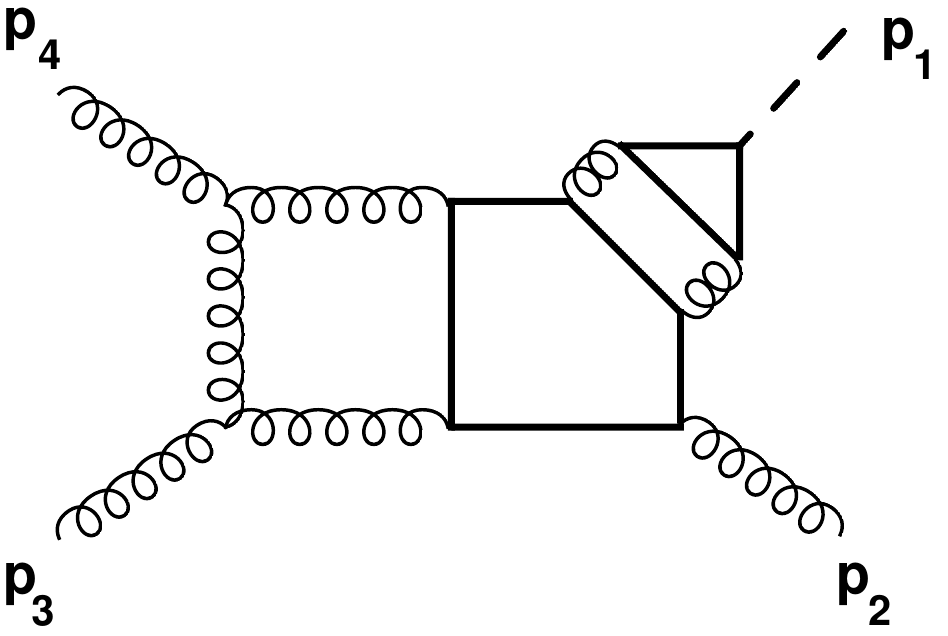} \end{subfigure} \\[1mm]
\begin{subfigure}{0.19\textwidth} \includegraphics[scale=0.19]{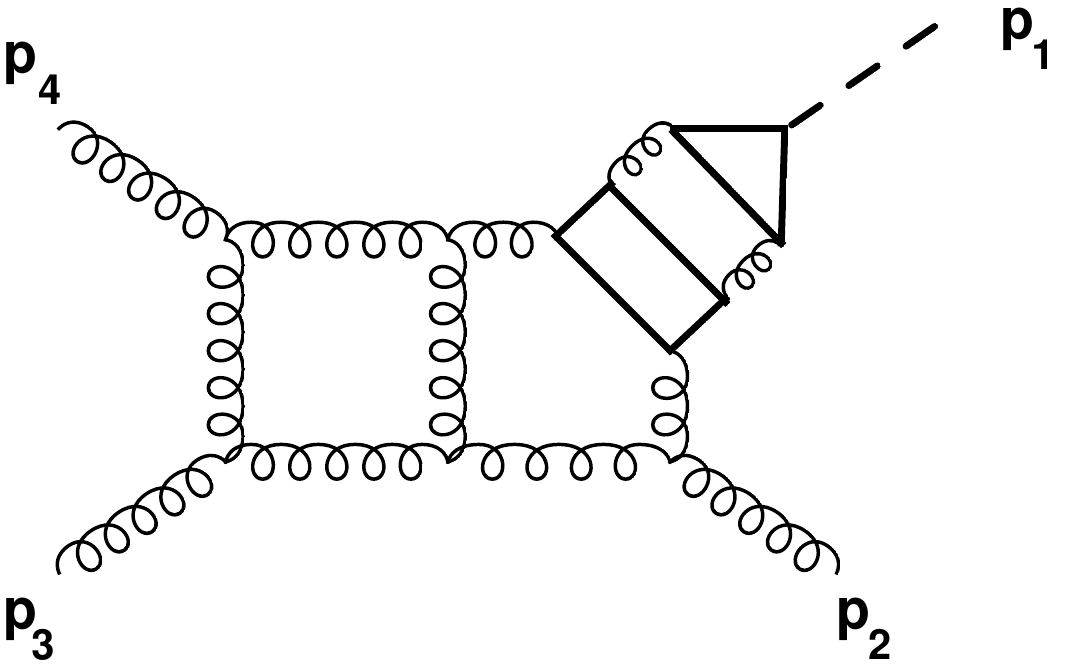} \end{subfigure}
\begin{subfigure}{0.19\textwidth} \includegraphics[scale=0.14]{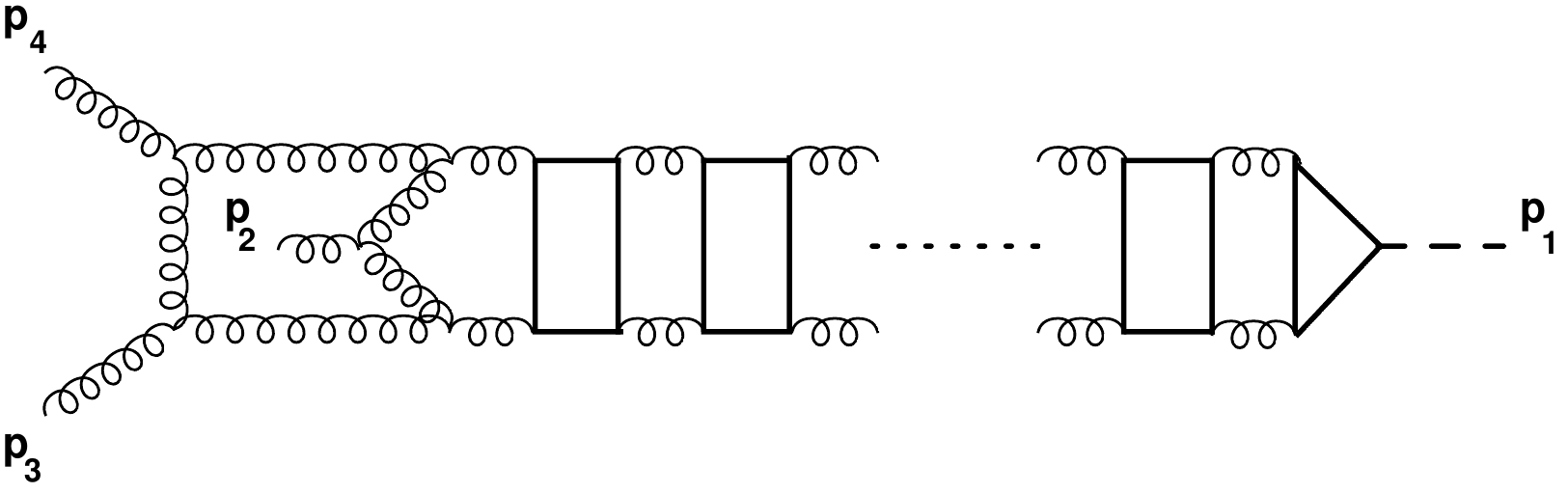} \end{subfigure}
\begin{subfigure}{0.19\textwidth} \includegraphics[scale=0.19]{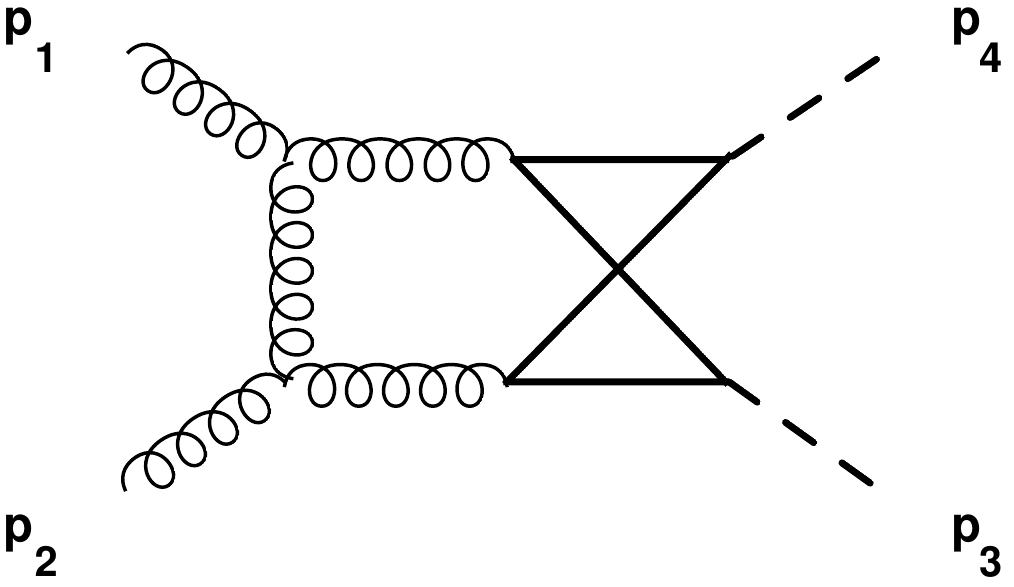} \end{subfigure}
\begin{subfigure}{0.19\textwidth} \includegraphics[scale=0.19]{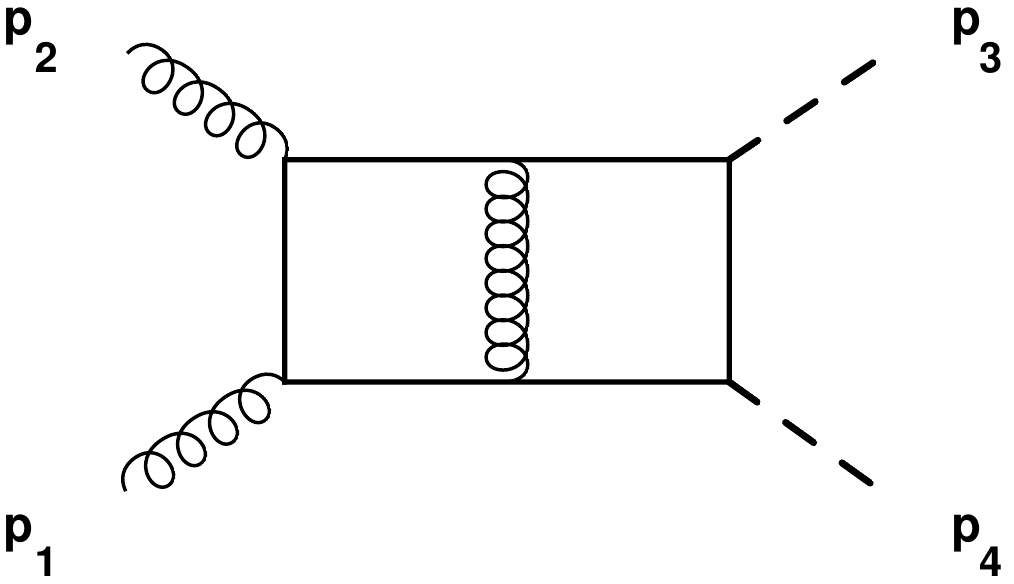} \end{subfigure}
\begin{subfigure}{0.19\textwidth} \includegraphics[scale=0.19]{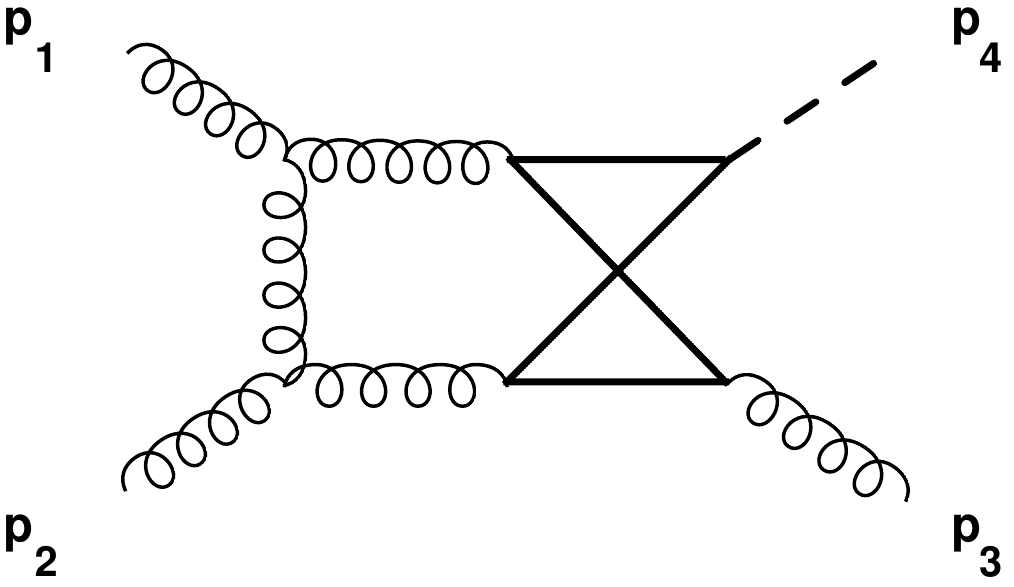} \end{subfigure} \\[1mm]
\begin{subfigure}{0.19\textwidth} \includegraphics[scale=0.19]{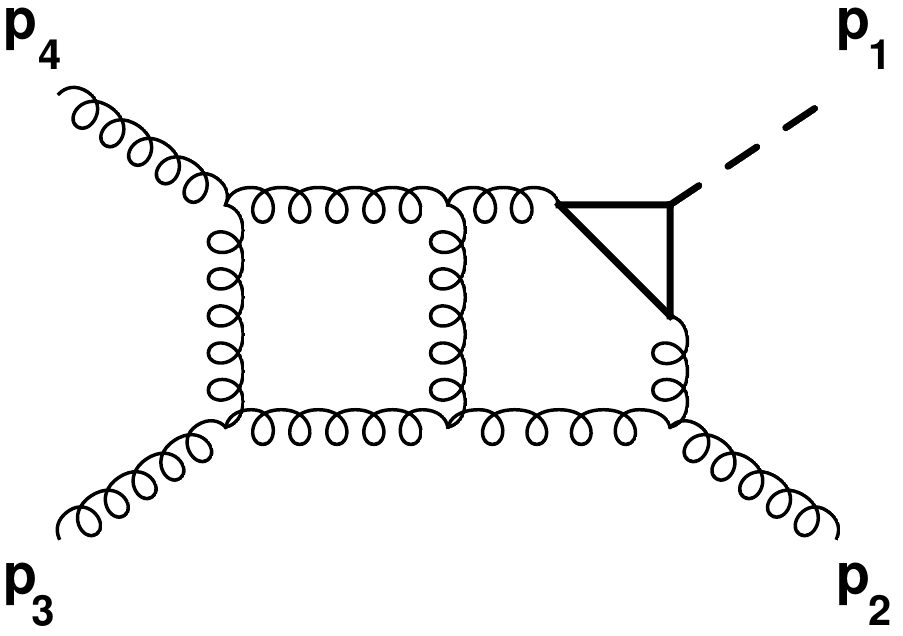} \end{subfigure}
\begin{subfigure}{0.19\textwidth} \includegraphics[scale=0.19]{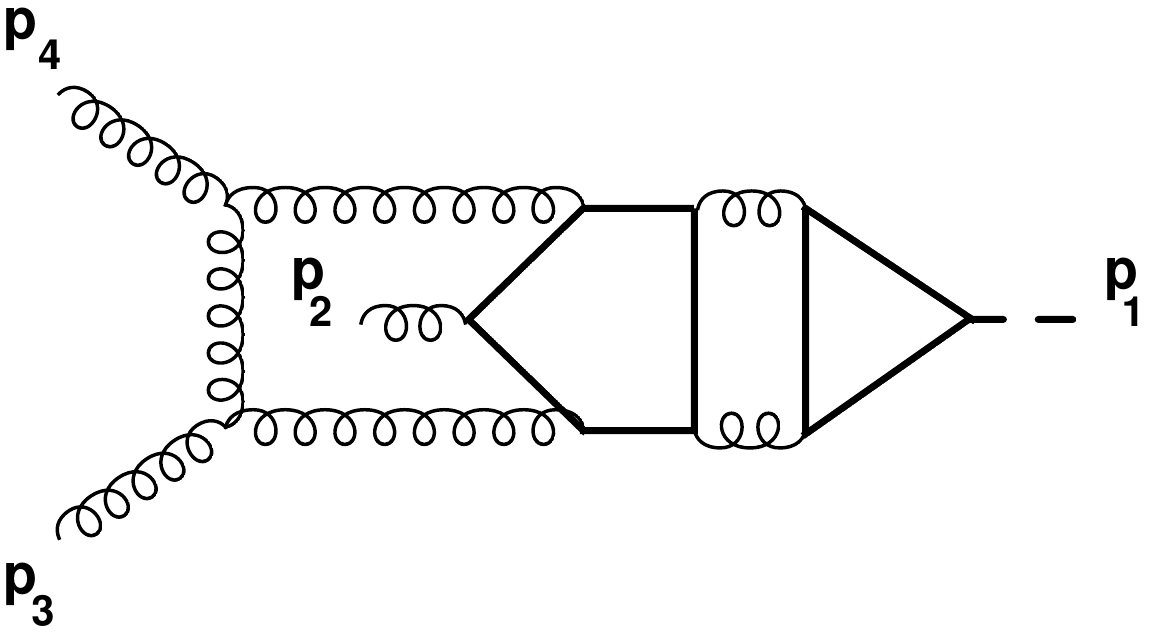} \end{subfigure}
\begin{subfigure}{0.19\textwidth} \includegraphics[scale=0.14]{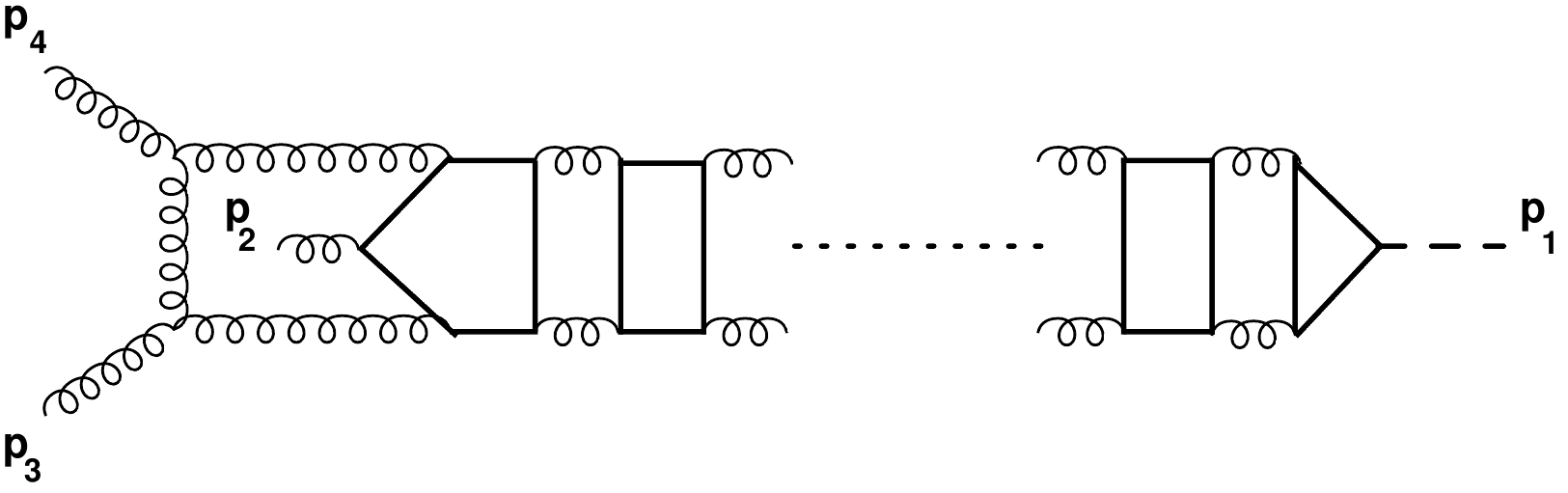} \end{subfigure}
\begin{subfigure}{0.19\textwidth} \includegraphics[scale=0.14]{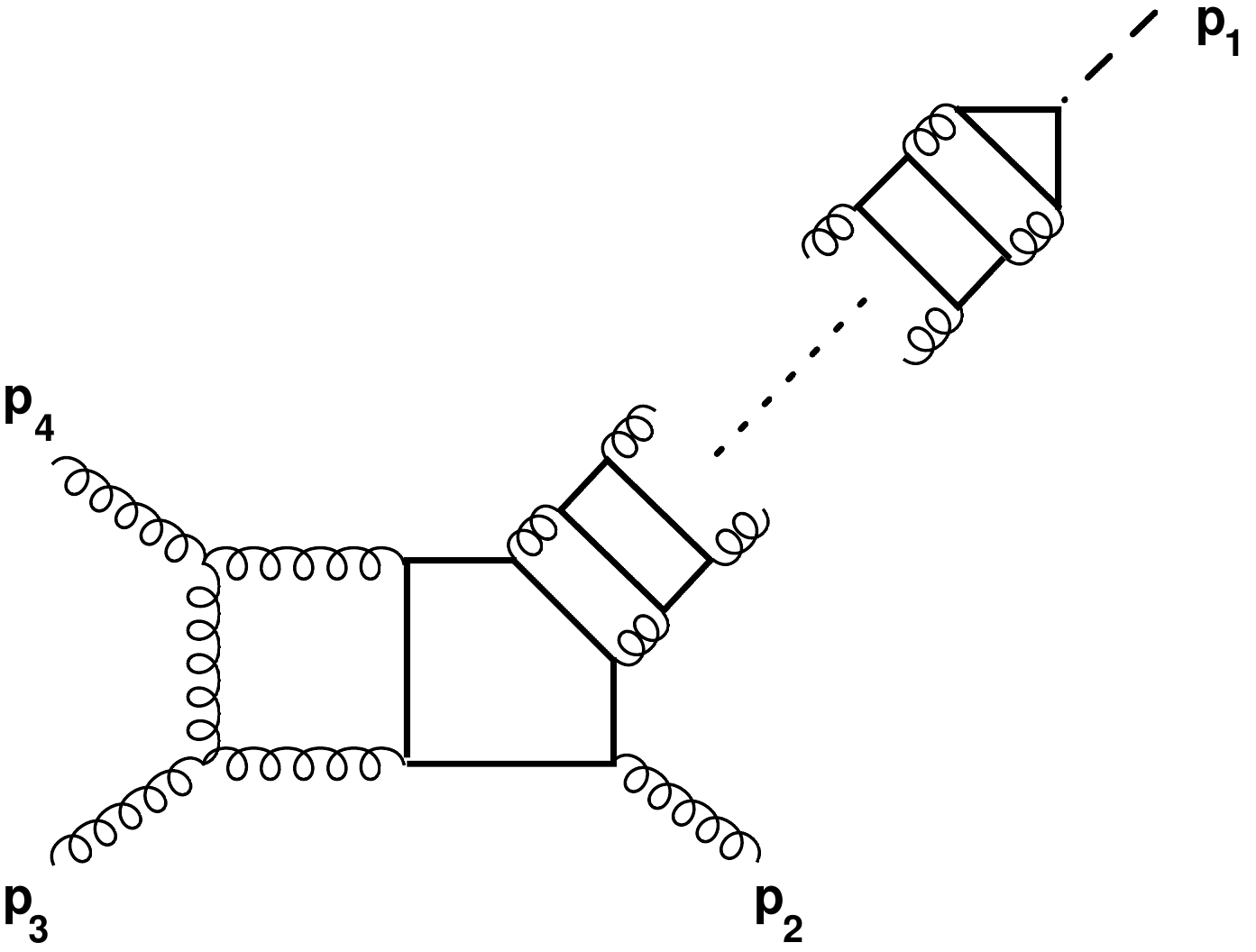} \end{subfigure}
\begin{subfigure}{0.19\textwidth} \includegraphics[scale=0.14]{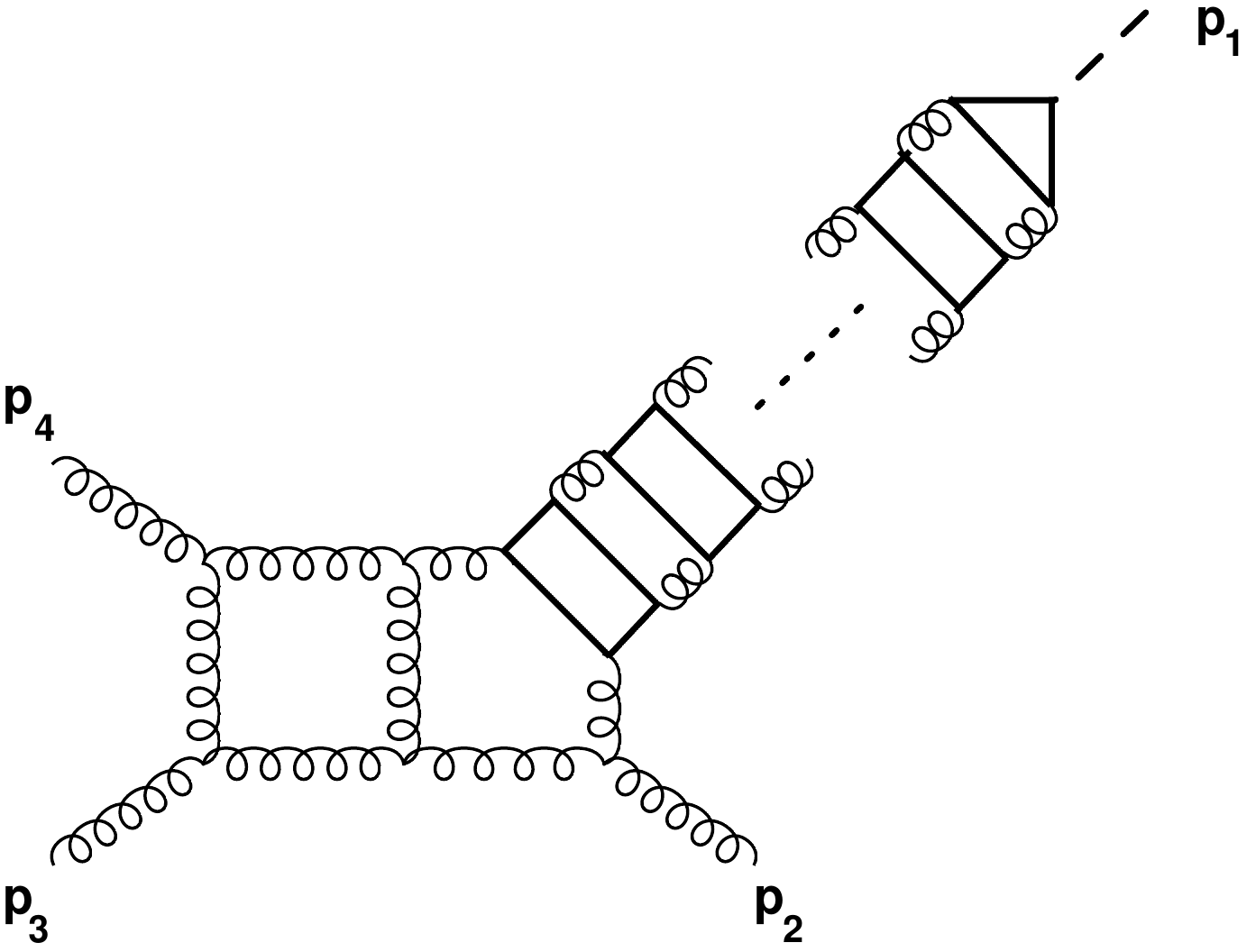} \end{subfigure} \\[1mm]
\begin{subfigure}{0.19\textwidth} \includegraphics[scale=0.19]{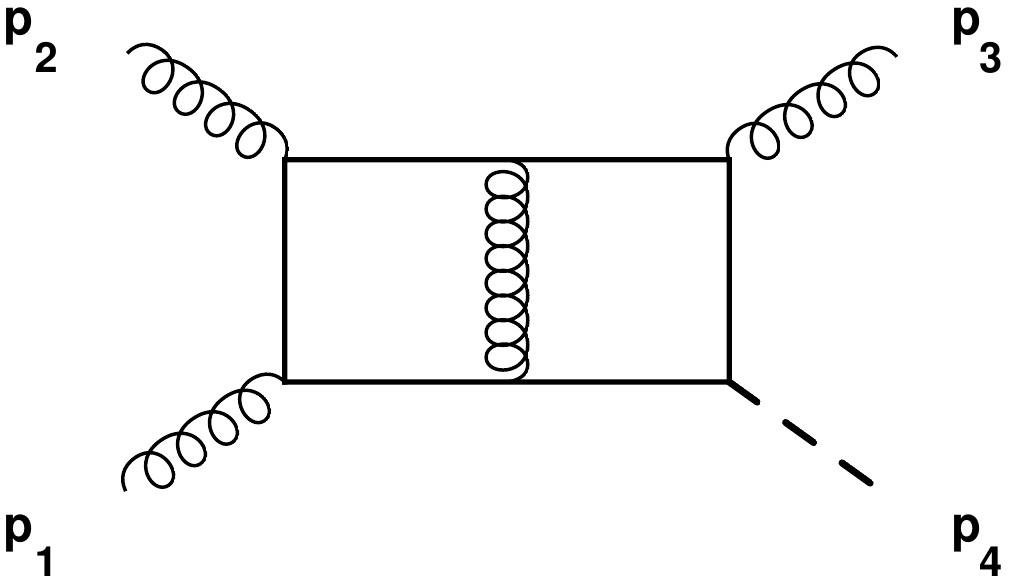} \end{subfigure}
\begin{subfigure}{0.19\textwidth} \includegraphics[scale=0.19]{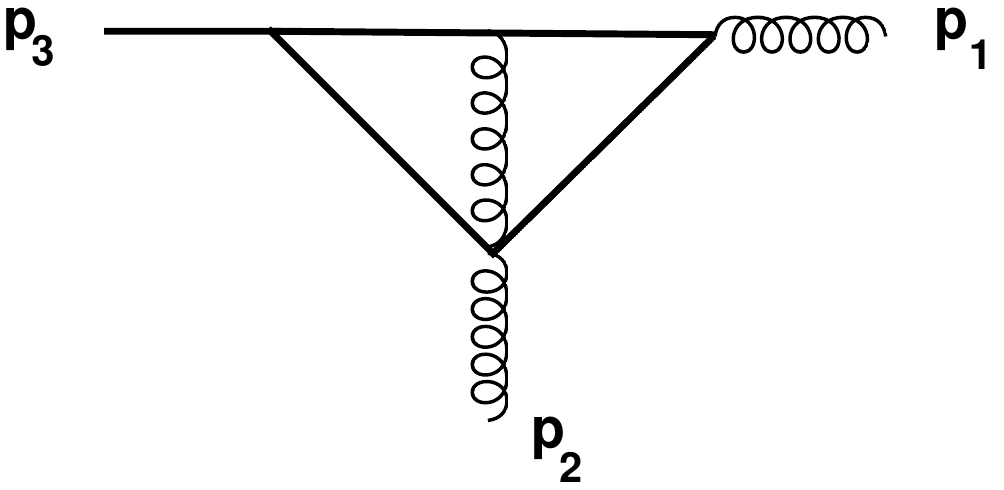} \end{subfigure}
\begin{subfigure}{0.19\textwidth} \hspace{0.5mm} \includegraphics[scale=0.19]{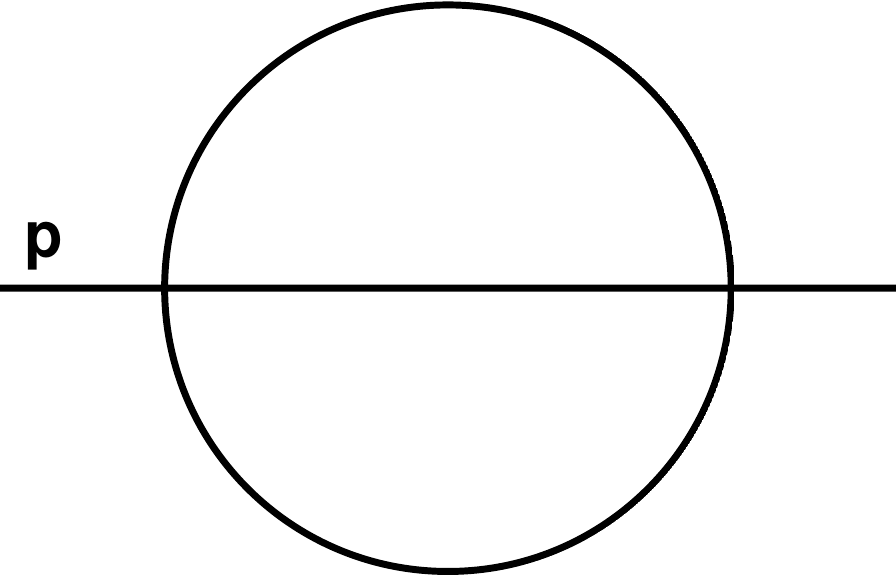} \end{subfigure}
\begin{subfigure}{0.19\textwidth} \includegraphics[scale=0.19]{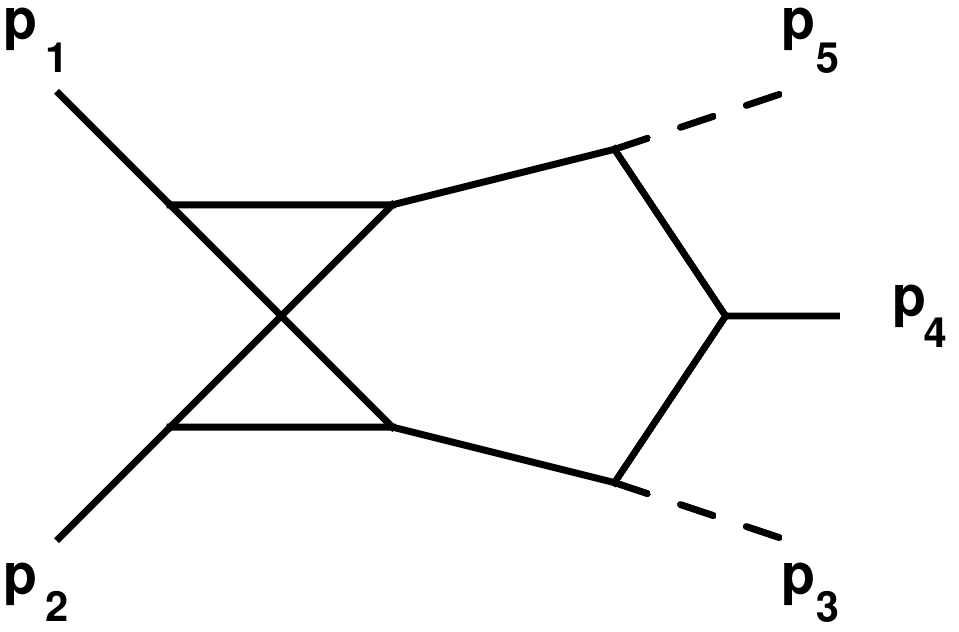} \end{subfigure}
\begin{subfigure}{0.19\textwidth} \includegraphics[scale=0.19]{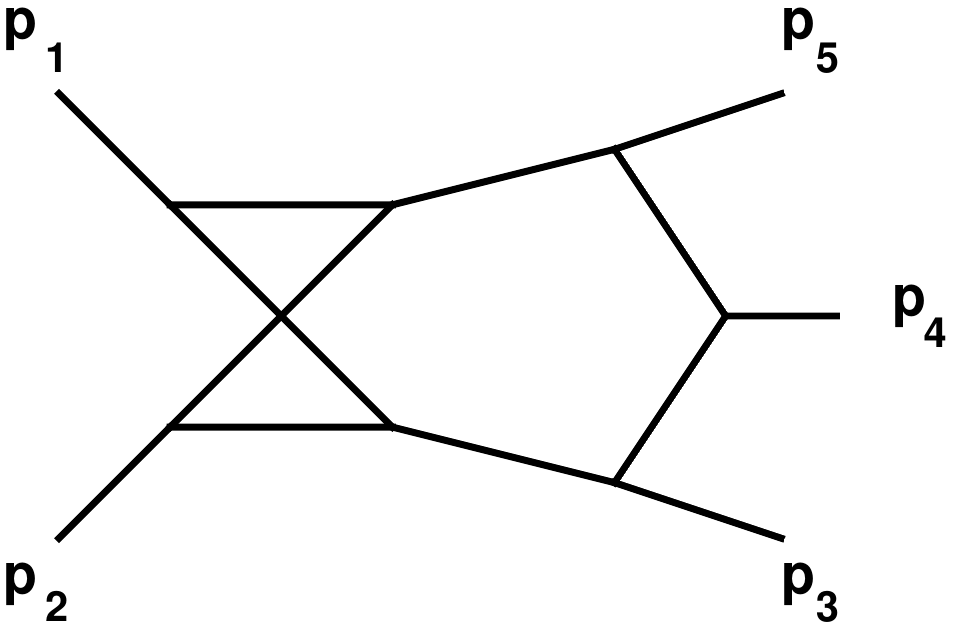} \end{subfigure} \\[1mm]
\begin{subfigure}{0.19\textwidth} \includegraphics[scale=0.19]{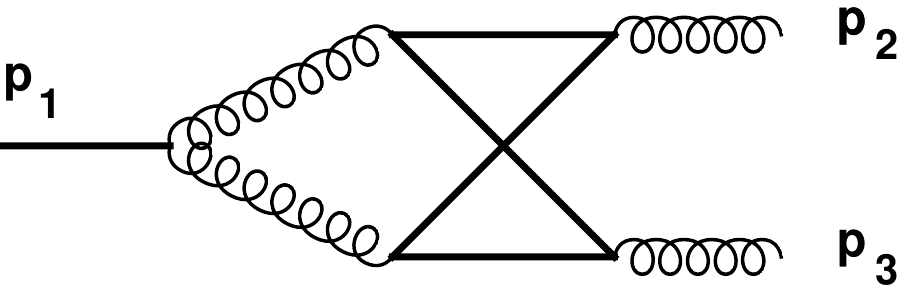} \end{subfigure}
\begin{subfigure}{0.19\textwidth} \includegraphics[scale=0.19]{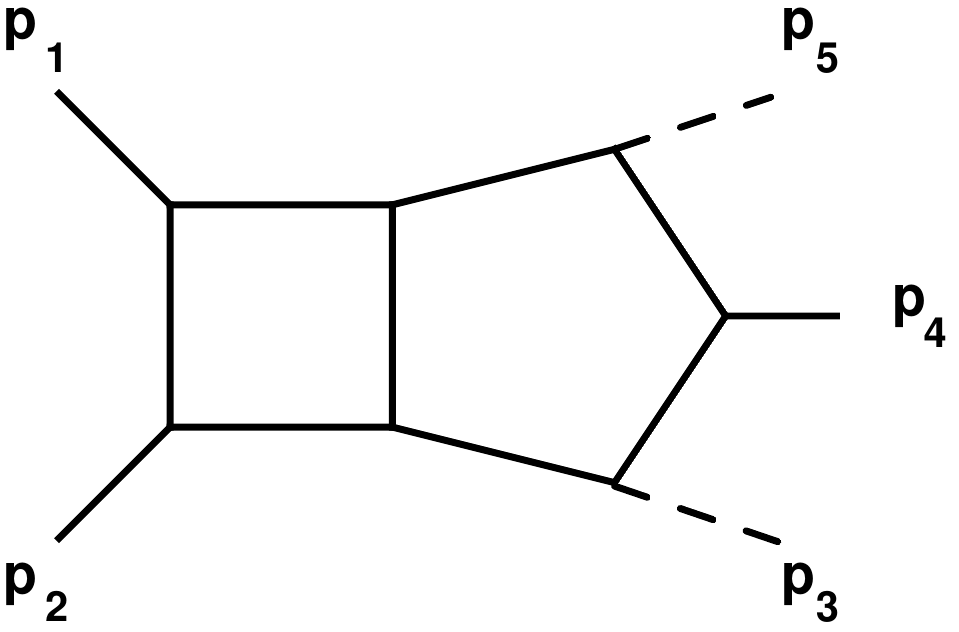} \end{subfigure}
\begin{subfigure}{0.19\textwidth} \includegraphics[scale=0.19]{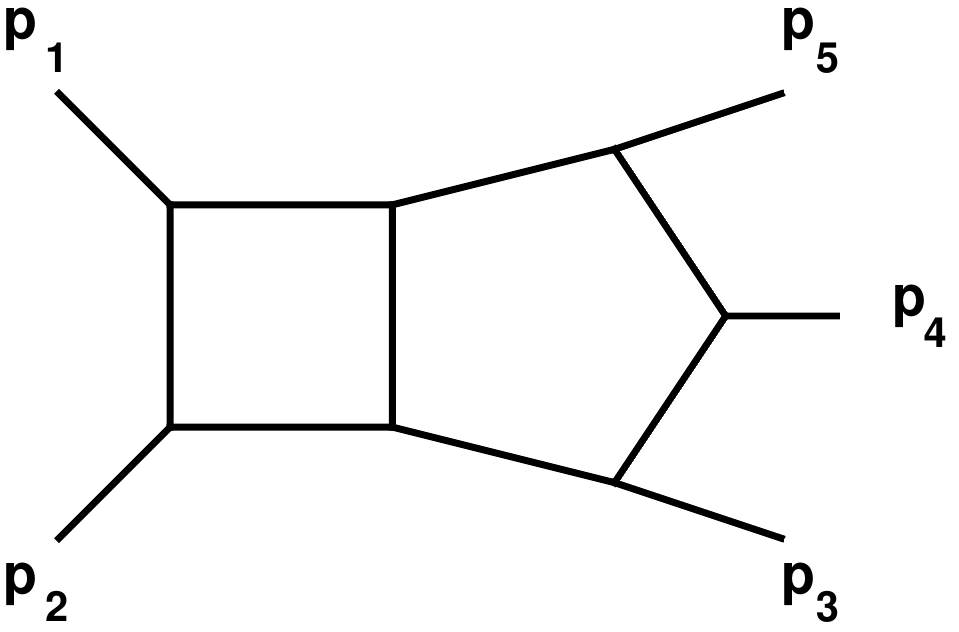} \end{subfigure}
\begin{subfigure}{0.19\textwidth} \includegraphics[scale=0.19]{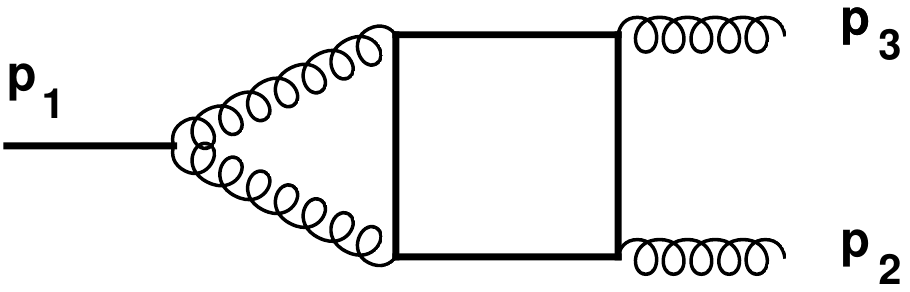} \end{subfigure}
\begin{subfigure}{0.19\textwidth} \includegraphics[scale=0.19]{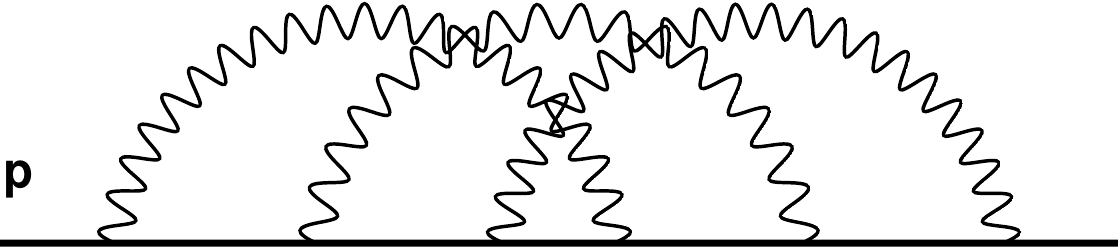} \end{subfigure}
\caption{Univariate examples from ref.~\cite{Frellesvig:2019kgj}.}
\label{fig:univariateall}
\end{figure}

With the same method we were able to get correct relations on the maximal cut for a large number of Feynman integrals, most of which are depicted in fig.~\ref{fig:univariateall}.

\section{Discussion}

In this work we elaborated on the relation between Feynman integrals and intersection theory, first presented in~\cite{Mastrolia:2018uzb} and further developed in~\cite{Frellesvig:2019kgj, Frellesvig:2019uqt, Frellesvig:2020qot}.\\
The focus of this contribution was on the univariate version of the theory, and for the multivariate generalization see~\cite{Frellesvig:2019uqt, Frellesvig:2020qot} or the contribution~\cite{ManojProceedings}.

We showed how Feynman integrals have a natural vector space structure with an inner product, allowing for new ways to compute their reduction onto a basis of master integrals.\\
In particular, we described the evaluation of univariate intersection numbers for twisted cocycles, which are the key ingredients of the master decomposition formula eq.~\eqref{eq:decomp}. 
We applied the master decomposition formula to derive contiguity relations between hypergeometric functions and, by means of the loop-by-loop Baikov representation (the properties of which are further outlined in this work), to a large number of Feynman integrals.\\
Interesting future studies would be to systematically apply intersection theory to integral parameterizations different from Baikov, as well as refining the algorithm for the computation of the intersection numbers, for instance in ways that remove the explicit solution of the differential equation of eq.~\eqref{eq:difeq}.
Additionally fruitful results might be obtained by combining these promising new techniques with finite field methods.\\

\acknowledgments

We would like to thank our collaborators Federico Gasparotto, Stefano Laporta, Manoj K. Mandal, Pierpaolo Mastrolia, and Sebastian Mizera. Additionally we would like to thank all the participants of the MathemAmplitudes 2019 conference for a great time and stimulating discussions.

The work of HF was part of the HiProLoop project funded by the European Union's Horizon 2020 research and innovation programme under the Marie Sk{\l}odowska-Curie grant agreement 74717. This project has received funding from the European Union’s Horizon 2020 research and innovation program under the Marie Sk{\l}odowska-Curie grant agreement No. 847523 ‘INTERACTIONS’. The work of HF is partially supported by a Carlsberg Foundation Reintegration Fellowship.

\bibliographystyle{JHEP}
\bibliography{biblio}

\providecommand{\href}[2]{#2}\begingroup\raggedright\begin{thebibliography}{10}

\bibitem{Mastrolia:2018uzb}
P.~Mastrolia and S.~Mizera, {\it {Feynman Integrals and Intersection Theory}},
  {\em JHEP} {\bf 02} (2019) 139, [\href{http://arxiv.org/abs/1810.03818}{{\tt
  arXiv:1810.03818}}].

\bibitem{Frellesvig:2019kgj}
H.~Frellesvig, F.~Gasparotto, S.~Laporta, M.~K. Mandal, P.~Mastrolia,
  L.~Mattiazzi, and S.~Mizera, {\it {Decomposition of Feynman Integrals on the
  Maximal Cut by Intersection Numbers}},  {\em JHEP} {\bf 05} (2019) 153,
  [\href{http://arxiv.org/abs/1901.11510}{{\tt arXiv:1901.11510}}].

\bibitem{Frellesvig:2019uqt}
H.~Frellesvig, F.~Gasparotto, M.~K. Mandal, P.~Mastrolia, L.~Mattiazzi, and
  S.~Mizera, {\it {Vector Space of Feynman Integrals and Multivariate
  Intersection Numbers}},  {\em Phys. Rev. Lett.} {\bf 123} (2019), no.~20
  201602, [\href{http://arxiv.org/abs/1907.02000}{{\tt arXiv:1907.02000}}].

\bibitem{Frellesvig:2020qot}
H.~Frellesvig, F.~Gasparotto, S.~Laporta, M.~K. Mandal, P.~Mastrolia,
  L.~Mattiazzi, and S.~Mizera, {\it {Decomposition of Feynman Integrals by
  Multivariate Intersection Numbers}},
  \href{http://arxiv.org/abs/2008.04823}{{\tt arXiv:2008.04823}}.

\bibitem{ManojProceedings}
F.~Gasparotto and M.~Mandal, {\it {On the Application of Intersection Theory to
  Feynman Integrals: The mutivariate case}},  in {\em {MathemAmplitudes 2019:
  Intersection Theory and Feynman Integrals}}, 2021.

\bibitem{PierpaoloProceedings}
P.~Mastrolia, {\it {Intersection Theory and Feynman Integrals - An Overview}},
  in {\em {MathemAmplitudes 2019: Intersection Theory and Feynman Integrals}},
  2021.

\bibitem{Mizera:2020wdt}
S.~Mizera, {\it {Status of Intersection Theory and Feynman Integrals}},  {\em
  PoS} {\bf MA2019} (2019) 016, [\href{http://arxiv.org/abs/2002.10476}{{\tt
  arXiv:2002.10476}}].

\bibitem{Weinzierl:2020gda}
S.~Weinzierl, {\it {Applications of intersection numbers in physics}},  in {\em
  {MathemAmplitudes 2019: Intersection Theory and Feynman Integrals}}, 11,
  2020.
\newblock \href{http://arxiv.org/abs/2011.02865}{{\tt arXiv:2011.02865}}.

\bibitem{Brown:2020rda}
F.~Brown and C.~Duhr, {\it {A double integral of dlog forms which is not
  polylogarithmic}},  6, 2020.
\newblock \href{http://arxiv.org/abs/2006.09413}{{\tt arXiv:2006.09413}}.

\bibitem{Vanhove:2020qtt}
P.~Vanhove and F.~Zerbini, {\it {Building blocks of closed and open string
  amplitudes}},  in {\em {MathemAmplitudes 2019: Intersection Theory and
  Feynman Integrals}}, 7, 2020.
\newblock \href{http://arxiv.org/abs/2007.08981}{{\tt arXiv:2007.08981}}.

\bibitem{Bendle:2020iim}
D.~Bendle, J.~Boehm, W.~Decker, A.~Georgoudis, F.-J. Pfreundt, M.~Rahn, and
  Y.~Zhang, {\it {Module Intersection for the Integration-by-Parts Reduction of
  Multi-Loop Feynman Integrals}},  in {\em {MathemAmplitudes 2019: Intersection
  Theory and Feynman Integrals}}, 10, 2020.
\newblock \href{http://arxiv.org/abs/2010.06895}{{\tt arXiv:2010.06895}}.

\bibitem{Chetyrkin:1981qh}
K.~G. Chetyrkin and F.~V. Tkachov, {\it {Integration by Parts: The Algorithm to
  Calculate beta Functions in 4 Loops}},  {\em Nucl. Phys.} {\bf B192} (1981)
  159--204.

\bibitem{Laporta:2001dd}
S.~Laporta, {\it {High precision calculation of multiloop Feynman integrals by
  difference equations}},  {\em Int. J. Mod. Phys.} {\bf A15} (2000)
  5087--5159, [\href{http://arxiv.org/abs/hep-ph/0102033}{{\tt
  hep-ph/0102033}}].

\bibitem{Mizera:2019vvs}
S.~Mizera and A.~Pokraka, {\it {From Infinity to Four Dimensions: Higher
  Residue Pairings and Feynman Integrals}},
  \href{http://arxiv.org/abs/1910.11852}{{\tt arXiv:1910.11852}}.

\bibitem{Weinzierl:2020xyy}
S.~Weinzierl, {\it {On the computation of intersection numbers for twisted
  cocycles}},  \href{http://arxiv.org/abs/2002.01930}{{\tt arXiv:2002.01930}}.

\bibitem{Chen:2020uyk}
J.~Chen, X.~Xu, and L.~L. Yang, {\it {Constructing Canonical Feynman Integrals
  with Intersection Theory}},  \href{http://arxiv.org/abs/2008.03045}{{\tt
  arXiv:2008.03045}}.

\bibitem{Baikov:1996iu}
P.~A. Baikov, {\it {Explicit solutions of the multiloop integral recurrence
  relations and its application}},  {\em Nucl. Instrum. Meth.} {\bf A389}
  (1997) 347--349, [\href{http://arxiv.org/abs/hep-ph/9611449}{{\tt
  hep-ph/9611449}}].

\bibitem{Lee:2010wea}
R.~N. Lee, {\it {Calculating multiloop integrals using dimensional recurrence
  relation and $D$-analyticity}},  {\em Nucl. Phys. Proc. Suppl.} {\bf 205-206}
  (2010) 135--140, [\href{http://arxiv.org/abs/1007.2256}{{\tt
  arXiv:1007.2256}}].

\bibitem{Grozin:2011mt}
A.~G. Grozin, {\it {Integration by parts: An Introduction}},  {\em Int. J. Mod.
  Phys.} {\bf A26} (2011) 2807--2854,
  [\href{http://arxiv.org/abs/1104.3993}{{\tt arXiv:1104.3993}}].

\bibitem{Larsen:2015ped}
K.~J. Larsen and Y.~Zhang, {\it {Integration-by-parts reductions from unitarity
  cuts and algebraic geometry}},  {\em Phys. Rev.} {\bf D93} (2016), no.~4
  041701, [\href{http://arxiv.org/abs/1511.01071}{{\tt arXiv:1511.01071}}].

\bibitem{Harley:2017qut}
M.~Harley, F.~Moriello, and R.~M. Schabinger, {\it {Baikov-Lee Representations
  Of Cut Feynman Integrals}},  {\em JHEP} {\bf 06} (2017) 049,
  [\href{http://arxiv.org/abs/1705.03478}{{\tt arXiv:1705.03478}}].

\bibitem{Bosma:2017ens}
J.~Bosma, M.~Sogaard, and Y.~Zhang, {\it {Maximal Cuts in Arbitrary
  Dimension}},  {\em JHEP} {\bf 08} (2017) 051,
  [\href{http://arxiv.org/abs/1704.04255}{{\tt arXiv:1704.04255}}].

\bibitem{Frellesvig:2017aai}
H.~Frellesvig and C.~G. Papadopoulos, {\it {Cuts of Feynman Integrals in Baikov
  representation}},  {\em JHEP} {\bf 04} (2017) 083,
  [\href{http://arxiv.org/abs/1701.07356}{{\tt arXiv:1701.07356}}].

\bibitem{Papadopoulos:2015jft}
C.~G. Papadopoulos, D.~Tommasini, and C.~Wever, {\it {The Pentabox Master
  Integrals with the Simplified Differential Equations approach}},  {\em JHEP}
  {\bf 04} (2016) 078, [\href{http://arxiv.org/abs/1511.09404}{{\tt
  arXiv:1511.09404}}].

\bibitem{Gehrmann:2015bfy}
T.~Gehrmann, J.~M. Henn, and N.~A. Lo~Presti, {\it {Analytic form of the
  two-loop planar five-gluon all-plus-helicity amplitude in QCD}},  {\em Phys.
  Rev. Lett.} {\bf 116} (2016), no.~6 062001,
  [\href{http://arxiv.org/abs/1511.05409}{{\tt arXiv:1511.05409}}]. [Erratum:
  Phys. Rev. Lett.116,no.18,189903(2016)].

\bibitem{Lee:2013hzt}
R.~N. Lee and A.~A. Pomeransky, {\it {Critical points and number of master
  integrals}},  {\em JHEP} {\bf 11} (2013) 165,
  [\href{http://arxiv.org/abs/1308.6676}{{\tt arXiv:1308.6676}}].

\bibitem{cho1995}
K.~Cho and K.~Matsumoto, {\it {Intersection theory for twisted cohomologies and
  twisted Riemann's period relations I}},  {\em Nagoya Math. J.} {\bf 139}
  (1995) 67--86.

\bibitem{matsumoto1998}
K.~Matsumoto, {\it Intersection numbers for logarithmic $k$-forms},  {\em Osaka
  J. Math.} {\bf 35} (1998), no.~4 873--893.

\bibitem{Keiji-MATSUMOTO2013367}
K.~Matsumoto, {\it {Monodromy and Pfaffian of Lauricella's $F_D$ in Terms of
  the Intersection Forms of Twisted (Co)homology Groups}},  {\em Kyushu Journal
  of Mathematics} {\bf 67} (2013), no.~2 367--387.

\bibitem{Smirnov:2014hma}
A.~V. Smirnov, {\it {FIRE5: a C++ implementation of Feynman Integral
  REduction}},  {\em Comput. Phys. Commun.} {\bf 189} (2015) 182--191,
  [\href{http://arxiv.org/abs/1408.2372}{{\tt arXiv:1408.2372}}].

\end{thebibliography}\endgroup

\end{document}